\documentclass[prl,preprint,superscriptaddress,showkeys,showpacs,
 preprintnumbers,amsmath,amssymb,aps,floatfix]{revtex4-1}
\usepackage{graphicx}
\usepackage{epsfig}
\usepackage{amsfonts}
\usepackage{amsmath}
\usepackage{amssymb}
\usepackage[sort&compress]{natbib}
\usepackage{dcolumn}
\usepackage{multirow}
\usepackage{bigstrut}
\usepackage{mathrsfs}
\usepackage{type1cm}
\usepackage{placeins}
\usepackage{rotating}
\usepackage{slashed}
\usepackage{subfigure}
\begin{document}

\preprint{\today}

\title{Limits on Lorentz violation from forbidden $\beta$ decays}

\author{J. P. Noordmans}
\author{H. W. Wilschut}
\author{R. G. E. Timmermans}

\affiliation{KVI, University of Groningen, Zernikelaan 25,
                   NL-9747 AA Groningen, The Netherlands}

\date{\today}
\vspace{3em}

\begin{abstract}
\noindent
Forbidden (slow) $\beta$ decays offer new opportunities to test the invariance of the weak interaction
under Lorentz transformations. Within a general effective field theory framework we analyze and reinterpret 
the only two relevant experiments, performed in the 1970s, dedicated to search for a preferred direction
in space in first- and second-forbidden $\beta$ decays. We show that the results of these experiments put
strong and unique limits on Lorentz violation, and in particular on the presence of several interactions
in the modern Lorentz-violating Standard Model extension. We discuss prospects to improve on these limits.
\end{abstract}
%\keywords{}
\maketitle

%\section{Introduction}
{\it Motivation\/} --
Nuclear $\beta$ decay has played a central role in the development of the electroweak sector of the
Standard Model (SM) of particle physics. The discovery of parity violation in the $\beta$ decay of
$^{60}$Co~\cite{Lee56,Wu57} led to the ``$V-A$'' theory of the weak interaction, and subsequently
to the understanding that $\beta$ decay is mediated by the $W$-boson that couples to left-handed
fermions. Present-day $\beta$-decay experiments search for deviations from the SM due to ``non-$V-A$''
currents, resulting for instance from the exchange of charged vector bosons that couple to right-handed
fermions or of charged scalar bosons~\cite{Her01,Sev11}.

In Ref.~\cite{Noo13a} we proposed that $\beta$-decay experiments offer interesting opportunities to
test the validity of Lorentz invariance of the weak interaction ({\em cf.} also Ref.~\cite{Dia13}). The search
for violations of Lorentz invariance is nowadays motivated by attempts to unify the SM with general
relativity~\cite{Lib09}. Some of these theories of ``quantum gravity'' predict Lorentz-violating signals that
could be detectable in high-precision experiments at low energy. The results of some recent searches
are reported in Refs.~\cite{Ben08,Alt09,Bro10,Smi11,Hoh13}. The experimental evidence for Lorentz and
in particular rotational invariance of the weak interaction, which has a bad track record for obeying
symmetries, however, is surprisingly poor, as pointed out already in Ref.~\cite{Lee56}.

In order to guide and interpret such $\beta$-decay experiments,  we developed a general approach
to the violation of Lorentz invariance in neutron and allowed nuclear $\beta$ decay~\cite{Noo13a}.
The effective Lorentz-violating Hamiltonian density is given by the current-current interaction
\begin{equation}
   \mathcal{H}_\beta  =  (g^{\mu\nu}+\chi^{\mu\nu})\left[\bar{\psi}_p(x)\gamma_\mu
                                               (C_V+C_A\gamma_5)\psi_n(x)\right] 
                                         \left[\bar{\psi}_e(x)\gamma_\nu(1-\gamma_5)\psi_\nu(x)\right]+{\rm H.c.} \ ,
\label{eq:hamiltonian}
\end{equation}
where $g^{\mu\nu}$ is the Minkowski metric and $\chi^{\mu\nu}$ is a complex, possibly
momentum-dependent, tensor that parametrizes Lorentz violation. $C_V$ and $C_A$ are the conventional
vector and axial-vector coupling constants and H.c. denotes Hermitian conjugation. As shown in
Ref.~\cite{Noo13a}, this approach includes a wide class of Lorentz-violating effects, such as contributions
from a modified $W$-boson propagator
$\left\langle W^{\mu+}W^{\nu-}\right\rangle = -i\,(g^{\mu\nu}+\chi^{\mu\nu})/M_W^2$,
but also contributions from a Lorentz-violating vertex $-i\gamma_{\nu}(g^{\mu\nu}+\chi^{\mu\nu})$.
Measurements of $\beta$-decay observables provide limits on the values of the components of
the tensor $\chi^{\mu\nu}$. Moreover, such limits can be translated into bounds on the parameters
of the Standard Model Extension (SME)~\cite{Col98,Kos11}, which provides the most general
effective field theory framework for the spontaneous breaking of Lorentz and CPT symmetry.

An experiment to test Lorentz violation in the Gamow-Teller $\beta$ decay of polarized $^{20}$Na
was recently performed at KVI, Groningen~\cite{Wil13,Mul13}. This is the first dedicated experiment
on allowed $\beta$ decay. However, several years before the $W$-boson was discovered and long
before searches for Lorentz violation became fashionable, two isolated experiments were performed
that searched for a ``preferred'' direction in space in first-forbidden $^{90}$Y $\beta$ decay~\cite{New76,Wie72}
and in first-forbidden $^{137}$Cs and second-forbidden $^{99}$Tc $\beta$ decays~\cite{Ull78}. The
hope was that such forbidden decays would be more sensitive to violations of rotational invariance,
{\it i.e.} angular-momentum conservation. We have revisited these experiments and interpreted them
within our effective field theory framework. For that reason, we have extended the approach of
Ref.~\cite{Noo13a} to forbidden $\beta$ decays. The technical details can be found in Ref.~\cite{Noo13b}.
In this Letter, we show that the experiments of Refs.~\cite{New76,Wie72,Ull78} provide strong and
unique bounds on Lorentz violation in the weak interaction, and in particular on previously
unconstrained parameters of the SME.

%\section{Forbidden $\beta$ decays}
%----------------------------------------------
{\it Forbidden $\beta$ decays} --
Since nuclear states are characterized by spin and parity, it is customary to expand the lepton
current in the $\beta$-decay matrix element in multipoles. Compared to the multipole expansion
of the photon field in the atomic case this expansion is complicated, because both vector
and axial-vector currents contribute and two relativistic particles are involved, for which only the
total angular momentum of each particle is a good quantum number. Moreover, the $\beta$ particle
moves in the Coulomb field of the daughter nucleus. The lowest-order terms in the multipole expansion
correspond to the allowed approximation, which amounts to evaluating the lepton current at $r=0$
and neglecting relativistic effects for the nucleus. This implies that neither of the leptons
carries off orbital angular momentum.

Higher-order terms in the expansion correspond to forbidden transitions~\cite{Kon66,Wei61},
which are suppressed by one or more of the following small dimensionless quantities: $pR$, where
$p$ is the lepton momentum and $R$ the nuclear radius (this corresponds to the ratio of the nuclear
radius and the de Broglie wavelength of the lepton), $v_N$, the velocity of the decaying nucleon
in units of $c$, and $\alpha Z$, the fine-structure constant times the charge of the daughter nucleus.
The lowest power of these quantities that appears in the amplitude determines the degree in
which the transition is forbidden. The transitions are classified by the nuclear-spin change
$\Delta I=|I_i-I_f|$ and relative parity $\pi_i\pi_f=\pm1$ (parity change {\em no} or {\em yes}),
where $I_i$, $\pi_i$ and $I_f$, $\pi_f$ are the spins and parities of the parent and daughter
nucleus, respectively. $n$-times forbidden transitions with $\Delta I =n+1$ are called unique.
Such unique forbidden transitions are advantageous, since they depend on only one nuclear matrix
element, which cancels in the asymmetries that quantify Lorentz violation.

In Ref.~\cite{Noo13b} we derived the multipole expansion for the Lorentz-violating $\beta$-decay
Hamiltonian of Eq.~(\ref{eq:hamiltonian}). Because the tensor $\chi^{\mu\nu}$ contracts the nucleon
and lepton currents in an unconventional way, the possibility arises that angular momentum is no
longer conserved in the transition. In particular, it is now possible that $\Delta I = J+1$ for $\chi^{0k}$
and $\chi^{k0}$ and that $\Delta I = J+2$ for $\chi^{km}$, where $J$ is the total angular momentum
of the leptons and Latin superscripts run over space indices. In contrast, rotational invariance implies that
$\Delta I \le J$. At the same time, however, the suppression of the transitions is still for the most part
determined by the angular momentum of the leptons. Due to this, the parts of $\chi^{\mu\nu}$ that
connect to the spin-dependent nucleon current ($\chi^{k0}$ and $\chi^{km}$) can be enhanced by
a factor $\alpha Z/pR$ with respect to the Lorentz-symmetric contributions. This enhancement factor
occurs only in transitions with $\Delta I\ge 2$, {\it i.e.} starting from unique first-forbidden transitions.

%\section{Analysis of the old experiments}
%--------------------------------------------------------
{\it Analysis of the old experiments\/} --
In Ref.~\cite{New76} the $\beta$-decay chain
$^{90}$Sr$(0^+,30.2\,a)$ $\rightarrow$ $^{90}$Y$(2^-,64.1\,h)$ $\rightarrow$ $^{90}$Zr$(0^+)$
was considered, wherein the $\beta^-$ decay of $^{90}$Y is a $\Delta I = 2$, {\em yes}, unique
first-forbidden transition. A search was made for dipole and quadrupole anisotropies in the angular
distribution of the electrons,
\begin{equation}
  W(\theta) = W_0\left( 1 + \varepsilon_1\cos\theta + \varepsilon_2\cos^2\theta \right) \ ,
   \label{eq:distribution}
\end{equation}
where $\theta$ is the angle between the electron momentum and a presumed preferred direction
in space. A 10 Ci $^{90}$Sr source was put in a vacuum chamber and the electron current it produced
was measured on a collector plate opposite the source, giving a solid angle of nearly 2$\pi$. The
source was made such that only high-energy electrons could come out, assuring that only the
current due to $^{90}$Y was measured. The endpoint of  $^{90}$Sr is too low to contribute significantly
to the current for this particular source. The chamber rotated about a vertical axis with a frequency of 0.75 Hz.
An anisotropy would result in a modulation of the detected current with a frequency of 0.75 or 1.5 Hz,
depending on the dipole or quadrupole nature of the anisotropy.

\begin{table}[t]
\centering
\setlength{\tabcolsep}{10pt}
\begin{tabular}{c|r|r|r}
\hline\hline
 Asymmetry $\delta$ &  $10^8\times a_0$ & $10^8\times a_1$ & $10^8\times a_2$ \\
\hline
  $N\!S$       & $-1.9\pm1.0$ & $3.2\pm1.9$ & $1.7\pm1.9$ \\
  $EW$       & $\;\;\:1.1\pm1.0$ & $2.9\pm1.9$ & $1.9\pm1.9$ \\
  2$\nu$    & $-1.0\pm1.0$ & $0.5\pm1.7$ & $0.7\pm1.7$ \\
\hline\hline
\end{tabular}
\caption{The measured values~\cite{New76} for $a_{0,1,2}$ of Eq.~(\ref{eq:delta}).}
\label{tab:limits}
\end{table}

The data were analyzed in terms of two dipole current asymmetries,
\begin{equation}
   \delta_{N\!S} = 2\,\frac{i_N-i_S}{i_N+i_S} \ , \;\;\;\;\;
   \delta_{EW} = 2\,\frac{i_E-i_W}{i_E+i_W} \ ,
\end{equation}
and one quadrupole asymmetry,
\begin{equation}
   \delta_{2\nu} = 2\,\frac{i_N+i_S-i_E-i_W}{i_N+i_S+i_E+i_W} \ ,
\end{equation}
where $N$, $S$, $E$, $W$ mean north, south, east, and west, and where for instance
$i_N$ denotes the mean current in the lab-fixed northern quadrant of the chamber's
rotation. These current asymmetries $\delta$ were fitted as functions of sidereal time as
\begin{equation}
  \delta = a_0 + a_1\sin(\omega t+\phi_1) + a_2\sin(2\omega t+\phi_2) \ ,
  \label{eq:delta}
\end{equation}
where $\omega$ is the angular rotation frequency of the Earth. The extracted coefficients
$a_{0,1,2}$ are given in Table~\ref{tab:limits}. Relative phases between the different
asymmetries were not considered and the phases $\phi_{1,2}$ between the amplitudes
were not reported. Such relations would have provided stronger constraints on Lorentz violation.

By using Eq.~\eqref{eq:distribution} the  expressions for $a_{0,1,2}$ were determined. Scattering
inside the source, due to which the emission direction of the electrons gets partly lost when they leave
the sample, had to be taken into account. With a Monte-Carlo program the probability distribution
to detect an electron was determined, depending on the angle of its original direction with respect
to the normal of the source. This probability distribution was then folded with Eq.~\eqref{eq:distribution}.
The result for the current as function of the angle $\theta_n$ between the direction of the collector
plate and the presumed asymmetry axis reads~\cite{Corr}
\begin{equation}
   I(\theta) = I_0 \left[ 1+ \frac{C_1}{3+C_2}\,\varepsilon_1\cos\theta_n
                                      +  \frac{C_2}{15+5C_2}\,\varepsilon_2\cos2\theta_n \right] \ ,
\label{eq:probdistribution}
\end{equation}
with $C_1=1.26$ and $C_2=0.39$. After transforming this equation to a standard
Sun-centered reference frame, the upper limits $|\varepsilon_1| < 1.6\!\times\!10^{-7}$ and
$|\varepsilon_2| < 2.0\!\times\!10^{-6}$ were determined at 90\% confidence level (C.L.)~\cite{New76}.

We interpret the data in Table~\ref{tab:limits} by using the new formalism~\cite{Noo13a}, in 
which the differential decay rate for a unique first-forbidden transition is given by~\cite{Noo13b}
\begin{equation}
   \frac{dW}{d\Omega dE} \propto p^2 + q^2 + p^2 \,
      \frac{\alpha Z}{pR} \left[\frac{3}{10}\frac{p}{E}\left(\chi_r^{ij}\hat{p}^i\hat{p}^j-\tfrac{1}{3}\chi_r^{00}\right)
      - \frac{1}{2}\tilde{\chi}_i^l \hat{p}^l +  \chi_r^{l0}\hat{p}^l\right] \ ,
\label{decayrspch2}
\end{equation}
where $p$ and $E$ are the electron momentum and energy and $q=E_0-E$ is the neutrino momentum,
with $E_0$ the energy available in the decay. The proportionality factor contains phase space and one
``$\beta$ moment''~\cite{Kon66}, a matrix element that depends on the nuclear structure.
The subscripts $r$ and $i$ on the Lorentz-violating tensor indicate the real and imaginary part of $\chi^{\mu\nu}$,
respectively, and $\tilde{\chi}^l=\epsilon^{lmk}\chi^{mk}$. The Lorentz-invariant part of the decay rate has
the typical unique first-forbidden spectrum shape $\sim p^2+q^2$. The Lorentz-violating part scales with
$\alpha Z/pR$, which proves that forbidden transitions can be more sensitive to angular-momentum violation,
compared to allowed ones. The enhancement is about one order of magnitude for a typical transition.

\begin{table*}[tb]
\centering
\begin{tabular}{c|c|c|c}
\hline\hline
 Asymmetry $\delta$ &  $a_0$  &  $a_1$ &  $a_2$  \\
\hline
  $N\!S$    & $-1.3\left[2X_r^{30}-\tilde{X}_i^3\right]$ &
  $1.4\left[\left(2X_r^{20}-\tilde{X}_i^2\right)^2 + \left(2X_r^{10}-\tilde{X}_i^1\right)^2\right]^{1/2}$ & $0$ \\
  $EW$     & $-0.63\left[2X_r^{30}-\tilde{X}_i^3\right]$ &
  $1.8\left[\left(2X_r^{20}-\tilde{X}_i^2\right)^2 + \left(2X_r^{10}-\tilde{X}_i^1\right)^2\right]^{1/2}$ & $0$ \\
  2$\nu$    & $0.0090\left[3X_r^{33}-X_r^{00}\right]$ &
  $0.031\left[\left(X_r^{13}+X_r^{31}\right)^2 + \left(X_r^{23}+X_r^{32}\right)^2\right]^{1/2}$ &
  $0.033\left[\left(X_r^{12}+X_r^{21}\right)^2 \right.%+ \left(X_r^{22}-X_r^{11}\right)^2\right]^{1/2}$ \\
  $ \\
 & & & \multicolumn{1}{r}{$\left.+\left(X_r^{22}-X_r^{11}\right)^2\right]^{1/2}$}\\
\hline\hline
\end{tabular}
\caption{The theoretical predictions from Eq.~\eqref{decayrspch2} for $a_{0,1,2}$ of Eq.~(\ref{eq:delta}).}
\label{tab:theory}
\end{table*}

With Eq.~\eqref{decayrspch2} instead of Eq.~\eqref{eq:distribution}, the remaining part of the analysis
parallels the analysis of Ref.~\cite{New76} summarized above. It requires a simulation of the
electron trajectories with the modified weight of the Lorentz-violating part of the expression, which, however,
would entail a small modification of the original simulation. Therefore, instead, we integrate the Lorentz-violating
part of Eq.~\eqref{decayrspch2} over the energy of the detected electrons, including the energy-dependent
phase-space factor $\propto pEq^2 F(Z,E)$, with $F(Z,E)$ the Fermi function. We integrate over the top
23.4\% of the energy spectrum, since the detector covered a $2\pi$ solid angle and 11.7\% of the electrons
escaped the source and were collected~\cite{New76}. With this simplified procedure we may do the angular
folding of Eq.~\eqref{decayrspch2} with the original detection probability distribution. We checked that
the limits derived below are not affected by more than 4\% when changing the integrated fraction of
the energy spectrum by a factor two.

We transform the tensor $\chi^{\mu\nu}$, defined in the laboratory frame, to the tensor $X^{\mu\nu}$ defined
in the Sun-centered frame~\cite{Kos11}, by using $\chi^{\mu\nu} = R^{\mu}_{\;\;\rho}R^{\nu}_{\;\;\sigma}X^{\rho\sigma}$,
with $R$ the apropriate rotation matrix~\cite{Noo13a}. In this way, we obtain the theoretical expressions for
the coefficients $a_{0,1,2}$ listed in Table~\ref{tab:theory}. The numerical factors in front of the coefficients in
Table~\ref{tab:theory} are determined by the location of the experiment on Earth (the colatitude of New York
is about $49^\circ$), the constants $C_1$ and $C_2$, and two phase shifts in the amplifier used in the
experiment~\cite{New76}. The phase correlation between $\delta_{N\!S}$ and $\delta_{EW}$ that our
theory predicts was not measured. Therefore, $2X_r^{10}-\tilde{X}_i^1$ and $2X_r^{20}-\tilde{X}_i^2$
cannot be extracted separately and only the combined value can be determined. The phase shifts of the
amplifier are the reason that $a_0 \neq 0$ for $\delta_{EW}$. 

Comparing the experimental values in Table~\ref{tab:limits} to the theoretical predictions in
Table~\ref{tab:theory}, we derive the following limits on the Lorentz-violating coefficients at 95\% C.L.:
\begin{subequations}
\begin{eqnarray}
-6\!\times\!10^{-9}<2X_r^{30}-\tilde{X}_i^3 & < & 2\!\times\!10^{-8}\ , \label{o1} \\
-3\!\times\!10^{-6}<3 X_r^{33}- X_r^{00} & < & 1\!\times\!10^{-6}\ , \label{o2} \\
\left[(2X_r^{20}-\tilde{X_i^2})^2+(2X_r^{10}-\tilde{X}_i^{1})^2\right]^{1/2} & < & 4\!\times\!10^{-8} \ , \label{c1} \\
\left[(X_r^{13}+X_r^{31})^2 + (X_r^{23}+X_r^{32})^2\right]^{1/2} & < & 1\!\times\!10^{-6}\ , \\
\left[(X_r^{12}+X_r^{21})^2 + (X_r^{22}-X_r^{11})^2\right]^{1/2} & < & 1\!\times\!10^{-6}\ . \label{c2}
\end{eqnarray}
\label{bounds}
\end{subequations}
The limit of Eq.~(\ref{c1}) was obtained from $\delta_{EW}$ only, because of the phase correlation
between $\delta_{N\!S}$ and $\delta_{EW}$.

In Ref.~\cite{Ull78} a similar search was made in the $\Delta I = 2$, {\em yes}, unique first-forbidden
$\beta^-$ decay $^{137}$Cs$(\frac{7}{2}^+,30.2\,a)$ $\rightarrow$ $^{137m}$Ba$(\frac{11}{2}^-)$,
and in the $\Delta I = 2$, {\em no}, second-forbidden $\beta^-$ decay
$^{99}$Tc$(\frac{9}{2}^+,2.1\times10^5\,a)$ $\rightarrow$
$^{99}$Ru$(\frac{5}{2}^+)$, by looking for a modulation of the counting rate as a function of sidereal time.
An upper limit for a $\cos\theta$ or $\cos2\theta$ term of $3\!\times\!10^{-5}$ was found at 90\% C.L.
Although the accuracy was less than in the $^{90}$Y experiment, the setup in this experiment
had a higher angular resolution. For $^{137}$Cs decay Eq.~\eqref{decayrspch2} also applies,
while the expression for $^{99}$Tc decay is given by~\cite{Noo13b}
\begin{equation}
\frac{dW}{d\Omega dE} \propto \mathcal{M}_{3/2}^ 2 p^2 + \mathcal{M}_{1/2}^ 2 q^2
    + \ \mathcal{M}_{3/2}\mathcal{W}\,p^2\,\frac{\alpha Z}{pR}
    \left[\frac{3}{10}\frac{p}{E}\left(\chi_r^{ij}\hat{p}^i\hat{p}^j-\tfrac{1}{3}\chi_r^{00}\right)
    - \frac{1}{2}\tilde{\chi}_i^l \hat{p}^l +  \chi_r^{l0}\hat{p}^l\right] \ ,
\label{decayrspch2b}
\end{equation}
where $\mathcal{M}_{1/2}$ and $\mathcal{M}_{3/2}$ depend on three $\beta$ moments~\cite{Kon66} and
$\mathcal{W}=2\mathcal{M}_{3/2}-\mathcal{M}_{1/2}$. We use $\mathcal{M}_{3/2}/\mathcal{M}_{1/2}=0.735$,
such that the spectrum shape $\sim 0.54\,p^2+q^2$~\cite{Lip66,Rei74}.

The experimental setup in Ref.~\cite{Ull78} was made to measure electrons in two directions. In one direction
perpendicular to the Earth's rotation axis a count rate $N_S$ was measured, and in the direction parallel to the
Earth's rotation axis a count rate $N_P$. The observable $\mathcal{A}=N_S/N_P-1$ was then inspected
for sidereal variations. By using the expressions for the Lorentz-violating decay rates of $^{137}$Cs and
$^{99}$Tc, given in Eqs.~\eqref{decayrspch2} and \eqref{decayrspch2b}, we obtain an expression similar
to Eq.~(\ref{eq:delta}) for the observable $\mathcal{A}$. The amplitudes are proportional to the same
combinations of Lorentz-violating coefficients as found previously for $^{90}$Y.
Here $a_0$ is a combination of the terms found in the first column of Table~\ref{tab:theory}. The terms for $a_1$
can be separated to obtain the individual terms $2X_r^{10}-\tilde{X}_i^1$ and $2X_r^{20}-\tilde{X}_i^2$. Similarly
this can be done for $a_2$. The proportionality constants are larger than for the $^{90}$Y experiment. In particular,
for the terms found previously in the quadrupole asymmetry the sensitivity of this setup is a factor 10 to 100 higher.
However, the statistical accuracy in this experiment is much lower and the improvements on the bounds of
Eq.~(\ref{bounds}) are insignificant. The best case is $|3X_r^{33}-X_r^{00}|<8\!\times\!10^{-6}$ at 95\% C.L.,
instead of the bound $3\!\times\!10^{-6}$ of Eq.~(\ref{o2}).

%\section{Discussion and conclusions}
%--------------------------------------------------
{\it Discussion and outlook\/} --
By using experiments on forbidden $\beta$ decay, we have set strong limits on Lorentz
violation in the weak interaction, in particular on the tensor $\chi^{\mu\nu}$ that modifies
the $W$-boson propagator. The general bounds of Eq.~\eqref{bounds} can be translated
into bounds on SME parameters~\cite{ColPhD,Cam12}, in terms of which~\cite{Noo13a}
\begin{equation}
   \chi^{\mu\nu} = - k_{\phi\phi}^{\mu\nu} - i\,k_{\phi W}^{\mu\nu}/2g \ ,
\end{equation}
when we assume that $\chi^{\mu\nu}$ is momentum independent~\cite{Noo13b}; $g$ is the
SU(2) electroweak coupling constant. Since $k_{\phi\phi}$ has a real symmetric component
$k_{\phi\phi}^S$ and an imaginary antisymmetric component $k_{\phi\phi}^A$, while $k_{\phi W}$
is real and antisymmetric, we derive at 95\% C.L. the bounds:
\begin{subequations}
\begin{eqnarray}
-5\times 10^{-9} < (k^S_{\phi\phi})^{ZT}, %&& \nonumber \\
  (k^A_{\phi\phi})^{Y\!X}, (k_{\phi W})^{Y\!X} &<& 1\times 10^{-8}\ , \\
-1\times 10^{-6} < (k^S_{\phi\phi})^{Z\!Z} &<& 4 \times 10^{-7}\ , \\
-1\times 10^{-6} < (k^S_{\phi\phi})^{TT} &<& 3\times 10^{-6}\ , \\
 \left|(k^S_{\phi\phi})^{X\!X}\right|, \left|(k^S_{\phi\phi})^{YY}\right| &<& 1\times 10^{-6}\ , \\
 \left|(k^S_{\phi\phi})^{XT}\right|, \left|(k^S_{\phi\phi})^{YT}\right|, \left|(k^A_{\phi\phi})^{X\!Z}\right|, %&& \nonumber \\
 \left|(k^A_{\phi\phi})^{Y\!Z}\right|,
 \left|(k_{\phi W})^{X\!Z}\right|, \left|(k_{\phi W})^{Y\!Z}\right| &<& 2 \times 10^{-8}\ , \\
 \left|(k^S_{\phi\phi})^{XY}\right|, \left|(k^S_{\phi\phi})^{X\!Z}\right|, \left|(k^S_{\phi\phi})^{Y\!Z}\right| &<& 5\times 10^{-7} \ .
\end{eqnarray}
\label{smebounds}
\end{subequations}
We assumed that there are no cancellations between different parameters,  {\it i.e.} when deriving
a bound on one parameter, the others were set to zero. With that caveat, Eq.~(\ref{smebounds}) provides
the first strict direct bounds on these SME parameters in the electroweak sector. For the components
$\chi^{\mu\nu}_r+\chi^{\nu\mu}_r$, they improve recent bounds from pion decay~\cite{Alt13} by three
orders of magnitude. (Indirect bounds were previously obtained for some of these parameters~\cite{And04}.
The validity of these indirect bounds is addressed in Ref.~\cite{Alt13}.)

In order to improve on our bounds, a more sensitive $\beta$-decay experiment of the type performed
in Refs.~\cite{New76,Wie72,Ull78} could be designed. With theory input and by exploiting modern
detector systems a number of the drawbacks of these pioneering experiments can be overcome.
However, to reach their precision level will require long-running experiments with high-intensity sources.
Forbidden $\beta$ transitions with low $E_0$ are preferred, as seen from Eq.~(\ref{decayrspch2}) and
because of radiation safety.
We have shown that $\beta$ decays with a higher degree of forbiddenness do not further enhance
Lorentz violation~\cite{Noo13b}. To obtain direct bounds on all the components $\chi^{\mu\nu}_{r,i}$
one may consider measurements of allowed $\beta$ transitions~\cite{Noo13a}. Both the degrees of
freedom of the $\beta$ particle and of the parent spin can be used. The KVI experiment~\cite{Wil13,Mul13}
measures $\tilde{\chi}^l_i$ in $\beta$ decay of polarized nuclei. $\beta^-$ emitters that populate the
ground state of the daughter nucleus are preferred. Sources like $^{32,33}$P, $^{35}$S, $^{45}$Ca,
or $^{63}$Ni are excellent options. An alternative is to measure semileptonic decays of hadrons,
in particular when they are produced in high-energy facilities and decay in flight: The boost then
provides an enhancement of order $\gamma^2$, where $\gamma$ is the Lorentz factor. Finally,
orders of magnitudes could be gained at a future $\beta$-beam facility~\cite{Lin10}, where
high-intensity high-energy radioactive beams decay to produce electron-neutrino beams.

%\section{Acknowledgments}
{\it Acknowledgments\/} --
We are much indebted to professor Riley Newman for clarifying communications about his
experiment~\cite{New76} and for sending us Ref.~\cite{Wie72}. We acknowledge an exchange
with professor Jack Ullman about Ref.~\cite{Ull78}. This research was supported by the
Stichting voor Fundamenteel Onderzoek der Materie (FOM) under Programmes 104 and 114
and project 08PR2636.

\end{document}